\documentclass[pra,
superscriptaddress,
twocolumn,
aps,
10pt
]{revtex4-1}
\usepackage{amsmath,amssymb}
\usepackage{graphicx}
\usepackage{bbold}
\usepackage{braket}
\usepackage[utf8]{inputenc}
\usepackage{natbib}
\usepackage{color}

\renewcommand{\vec}[1]{\mathbf{#1}}

\newcommand{\eps}{\hat{\mathcal{E}}(\vec{r})}
\newcommand{\epsdag}{\hat{\mathcal{E}}^{\dagger}(\vec{r})}
\newcommand{\gs}{\hat{X}_{g}(\vec{r})}
\newcommand{\gsdag}{\hat{X}^{\dagger}_{g}(\vec{r})}

\newcommand{\gry}[1]{\hat{X}_{#1}}
\newcommand{\grydag}[1]{\hat{X}^{\dagger}_{#1}}
\newcommand*{\dprime}{\prime\prime}

\begin{document}

\title{Giant optical nonlinearities from Rydberg-excitons in semiconductor microcavities}

\author{Valentin Walther}
\affiliation{Department of Physics and Astronomy, Aarhus University, Ny Munkegade 120, DK 8000 Aarhus C, Denmark}
\affiliation{Max Planck Institute for the Physics of Complex Systems, N\"othnitzer Str. 38, 01187 Dresden, Germany}

\author{Robert Johne}
\affiliation{Max Planck Institute for the Physics of Complex Systems, N\"othnitzer Str. 38, 01187 Dresden, Germany}

\author{Thomas Pohl}
\affiliation{Department of Physics and Astronomy, Aarhus University, Ny Munkegade 120, DK 8000 Aarhus C, Denmark}
\affiliation{Max Planck Institute for the Physics of Complex Systems, N\"othnitzer Str. 38, 01187 Dresden, Germany}

\begin{abstract}
The realization of exciton-polaritons -- hybrid excitations of semiconductor quantum well excitons and cavity photons -- has been of great technological and scientific significance. In particular, the short-range collisional interaction between excitons has enabled explorations into a wealth of nonequilibrium and hydrodynamical effects that arise in weakly nonlinear polariton condensates. Yet, the ability to enhance optical nonlinearities would enable quantum photonics applications and open up a new realm of photonic many-body physics in a scalable and engineerable solid-state environment. Here we outline a route to such capabilities in cavity-coupled semiconductors by exploiting the giant interactions between excitons in Rydberg-states. We demonstrate that optical nonlinearities in such systems can be vastly enhanced by several orders of magnitude and induce nonlinear processes at the level of single photons.
\end{abstract}

\maketitle

The achievement of strong coupling between quantum-well excitons and optical photons in semiconductor microcavities \cite{weisbuch1992} has ushered in new lines of research on exciton-polariton systems. Their unique properties in combination with advanced semiconductor technology \cite{schneider2017} are exploited for the development of novel devices such as next-generation lasers \cite{schneider2013} but also offer a unique platform for fundamental studies of many-body phenomena \cite{carusotto2013}. Their excitonic component endows such polaritons with interactions that can drive a variety of collective phenomena from condensation \cite{Deng2002} and superfluidity \cite{amo2009} to solitons \cite{sich2012} and parametric amplification \cite{saba2001}. Yet strongly correlated states and nonlinear processes at the level of individual photons \cite{chang2014} are inherently difficult to realize due to the weak and short-range nature of typical exciton-exciton interactions \cite{carusotto2013,ferrier2011}.

\begin{figure}[t!]
\includegraphics[width=0.99\linewidth]{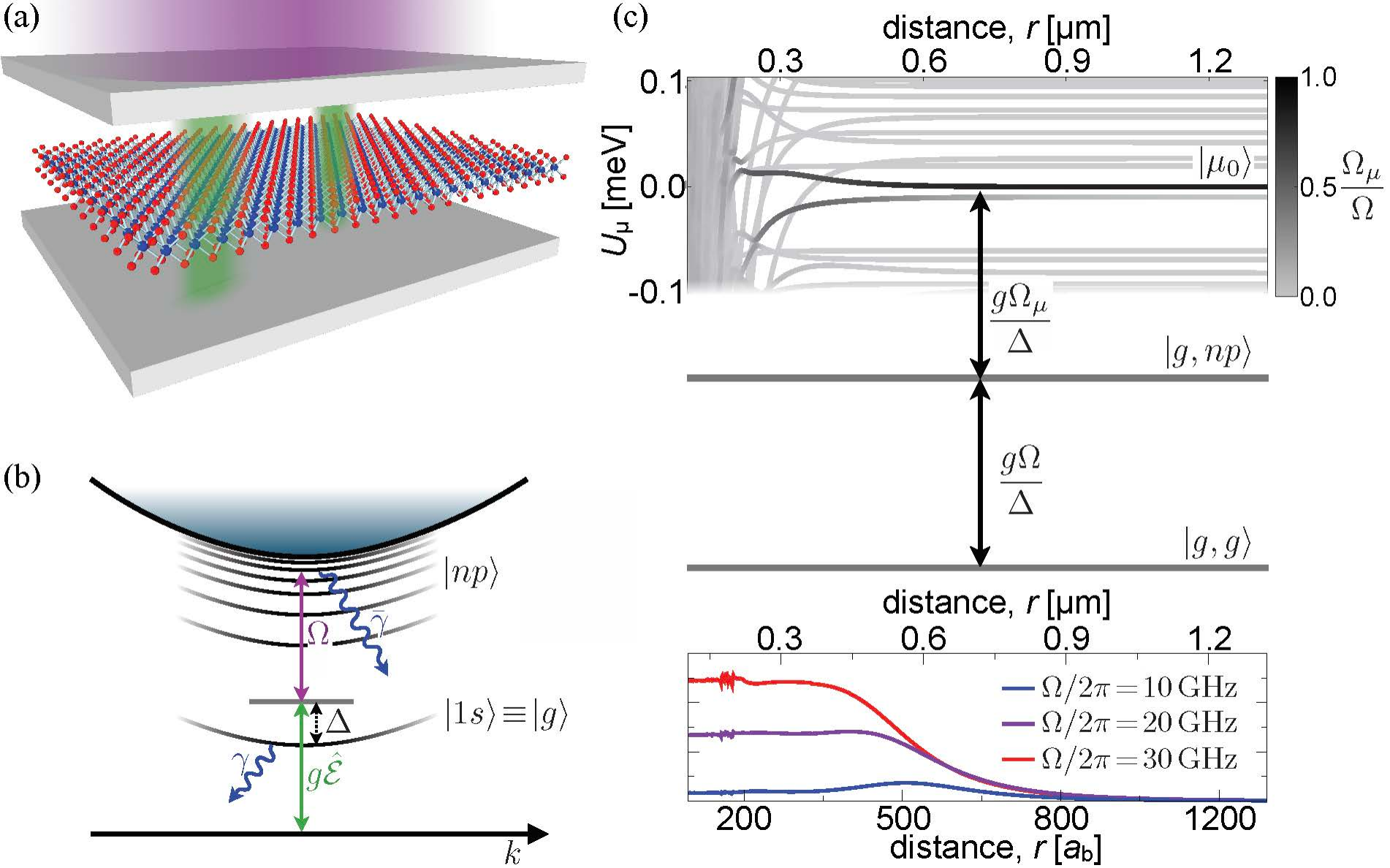}
\caption{(a) A Fabry-P\'erot cavity yields strong near resonant coupling of the cavity field $\hat{\mathcal{E}}$ (green) to a low-lying exciton state of a two-dimensional semiconductor. Another radiation field (purple) provides coupling to excitonic Rydberg states whose strong interactions generate a nonlinear response to the cavity field. 
(b) The cavity field generates deeply bound excitons with a coupling strength $g$ and single-photon detuning $\Delta$. The additional field provides for two-photon resonant Rydberg-state excitation with a Rabi frequency $\Omega$.
(c) The resulting two-photon coupling from singly ($|g, np\rangle$) to doubly ($|\mu\rangle$) excited  Rydberg states is strongly influenced by their interactions which cause significant potential energy shifts, $U_\mu(r)$, and distance-dependent coupling strengths $\Omega_\mu(r)$ (gray coloring), depicted around the $|\mu_0\rangle=|10p,10p\rangle$ pair state.
The resulting photon-photon interaction is shown in (d) for $g/2\pi=5561$Ghz \cite{liu2015}, $\Delta/2\pi = 700$Ghz, $\gamma/2\pi = 300$Ghz and $\bar{\gamma}/2\pi = 0.3$GHz.}
\label{fig1} 
\end{figure}

Here we describe how one can reach this quantum regime by dressing the photon field of semiconductor microcavities with strongly interacting Rydberg states of excitons. 
Excited states of excitons have been observed in transition metal dichalcogenide (TMDC) monolayers \cite{chernikov2014} and in Cuprous Oxide, where high lying Rydberg states with principal quantum numbers of up to $n=25$ could be demonstrated \cite{kazimierczuk2014}. Rydberg states feature a number of remarkable properties that are explored and exploited in atomic systems for cavity-QED experiments \cite{haroche2013}, quantum simulations  and information processing \cite{saffman2010} as well as quantum nonlinear optics \cite{firstenberg2016,murray2016}. Rydberg-exciton polaritons therefore offer a promising combination of such new capabilities afforded by the strong interactions between Rydberg states with the technological advantages of semiconductor photonics. Our theoretical framework permits to deduce the associated nonlinear optical response from the rather complex potential surfaces of interacting Rydberg-state manifolds (cf. Fig.\ref{fig1}c), and it indeed yields nonlinearities that exceed those of ground state systems \cite{ferrier2011} by many orders of magnitude. Remarkably, this vast enhancement can persist even in the presence of considerable Rydberg-state decoherence, up to $10^4$ times stronger then for corresponding atomic Rydberg states \cite{saffman2010}. This surprising behaviour is traced back to an exciton-blockade mechanism not previously discussed for atomic systems, and shown to permit the generation of strongly correlated quantum states of light at the level of individual photons. 

Figs.\ref{fig1}a and \ref{fig1}b illustrate the considered setup based on near-resonant generation of semiconductor excitons by the photon field inside an optical microcavity and their further excitation to a Rydberg state via strong externally applied radiation. Importantly, the resulting three-level driving scheme permits to establish approximate conditions of electromagnetically induced transparency \cite{fleischhauer2005} that enable strong light-matter coupling at greatly reduced photon losses.  
Its realization in semiconductors requires quasi-particle band gaps in the optical domain as well as large exciton binding energies. For example, $2p$ excitons of the yellow series in ${\rm Cu}_2{\rm O}$ \cite{kazimierczuk2014} are resonant with  $580$nm light and can be coupled to the strongly interacting $20s$ state at $50\mu$m. Semiconducting TMDC monolayers feature even larger exciton binding energies in the range of $0.3-0.7$ eV and a remarkably strong light-matter coupling \cite{wang2012}.
We calculate the non-hydrogenic excitonic Rydberg series of such single-layered TMDC materials \cite{glazov2017}, accounting for the screened  electron-hole interaction \cite{chernikov2014,Entin1972} as well as Berry curvature effects arising from the band structure topology \cite{srivastava2015,zhou2015} (cf. supplementary material, section \ref{sect:single_exciton}). In the following, we consider exciton states obtained for a binding energy of $0.3$eV, equal electron and hole masses of $0.26$m$_{\rm e}$ \cite{chernikov2014} and a Berry curvature of $0.15{\rm nm}^2$ \cite{zhou2015}. 
The latter gives rise to a distinct set of exciton states $|n,m\rangle$ at each the $K^{+}$- and the $K^{-}$-points of the Brillouin zone. These are characterized by the principal quantum number $n$ and the angular quantum number $m=0,\ldots, \pm(n-1)$.

The nature of the interaction between such high-$n$ excitons differs significantly from that of short-range collisions between low-lying exciton states \cite{carusotto2013}. Due to their enormous strength, the interaction between highly excited Rydberg-excitons typically becomes relevant at such large distances \cite{kazimierczuk2014} that exchange effects \cite{Shahnazaryan2016} become negligible and direct electrostatic interactions play the dominant role. 
Consequently, the interaction potential has to be determined non-perturbatively from the dipole-dipole coupling between different pair states, $|n_1m_1;n_2m_2\rangle$ (cf. supplementary material, section \ref{sect:dipole_interaction}). 
As a function of exciton-exciton separation $r$, this yields a set of molecular states $|\mu\rangle$, associated potential curves $U_\mu(R)$ and optical coupling strengths $\Omega_{\mu}(r)$ (cf. Fig.\ref{fig1}c).

\begin{figure}
\includegraphics[width=0.99\linewidth]{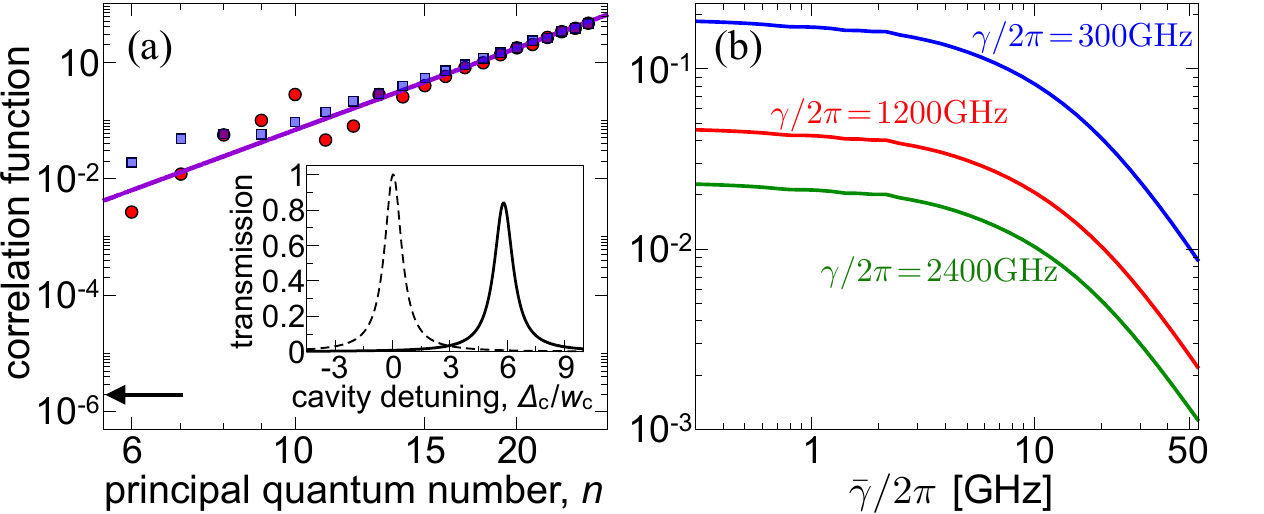}
\caption{(a) Integrated interaction strength $\alpha$ for the $m=1$ excitonic series at the $K^{+}$- (red) or $K^{-}$-point (blue) of the Brillouin zone with $\Omega/2\pi=60$GHz, $\bar{\gamma} = \gamma/n^{3}$ and the remaining parameters as in Fig. \ref{fig1}. 
The solid line shows the $\alpha\sim n^{14/3}$ scaling found for both $K$-points which results from the large-distance $\sim n^{11}$ dependence of the Rydberg-exciton interaction and the considered $\bar{\gamma}\sim n^{-3}$ decrease of their linewidth \cite{kazimierczuk2014}. For lower principal quantum numbers, however, the interaction strength features a significant valley-dependence. 
For comparison, the arrow indicates the nonlinearity measured \cite{ferrier2011} for ground-state excitons in GaAs. The inset shows corresponding transmission spectra upon changing the cavity frequency for $n=10$ in the linear regime (dashed line) and for small driving $|\mathcal{E}|^2 = 2.5 \times10^{-5}\mu m^{-2}$ (solid line). Both curves are scaled by the maximum of the linear transmission line. The cavity detuning $\Delta_{\rm c}$ is shown in units of the width $w_{\rm c}$ of the linear transmission line.
Panel (b) illustrates that the nonlinearity persists in the presence of significant Rydberg-state broadening, $\bar{\gamma}$, as shown for different values of $\gamma$ and $\Delta = 2 \gamma$ at constant $\frac{\Omega^2}{|\Gamma|} = 40$GHz.}
\label{fig2} 
\end{figure}

The ensuing consequences for the optical response are easiest understood within a simplified picture known from atomic systems \cite{saffman2010}, where one assumes $\bar{\gamma}=0$ and considers only a single pair state $|\mu_0\rangle$ with a van der Waals interaction potential $U_{\mu_0}\sim R^{-6}$ and negligible state mixing. By tuning the frequency of the Rydberg-excitation laser onto two-photon resonance one can establish EIT conditions within a frequency window $\Omega^2/|\Gamma|$ determined by the Rydberg-excitation Rabi frequency $\Omega$ and $\Gamma=\gamma+i2\Delta$ given by the decay rate $\gamma$ and the single-photon detuning $\Delta$ from the low-lying exciton line (see Fig.\ref{fig1}b). Therefore, the otherwise high optical susceptibility originating from the strong exciton-cavity coupling can be greatly reduced by resonant Rydberg-state coupling. This effect can be traced back to the formation of dark-state polaritons \cite{fleischhauer2005} whose coherence properties in the present case are predominantly limited by the Rydberg-state linewidth rather than the large decay rate of the low-lying exciton state.
The strong interactions between Rydberg states can, however, drastically modify this picture. 
In particular, the level shift, $U_{\mu_0}(R)$ induced by a single Rydberg excitation can be sufficient to inhibit further excitation in its vicinity and thereby expose the strong optical response of the low-lying transition. This interaction blockade, thus, provides a simple mechanism for the emergence of strong, spatially nonlocal optical nonlinearities \cite{rotschild2006} that has been demonstrated in a number of recent experiments in atomic systems \cite{firstenberg2016,murray2016}. 

In a semiconductor, however, the much larger energy scales for the light-matter coupling and stronger decoherence prompt the necessity of a more advanced theory that accounts for the collective coupling to a large manifold of strongly interacting Rydberg-exciton states (cf. supplementary material, section \ref{sect:hamiltonian}). To this end, we determine the nonlinearity for coherent light fields, described by the amplitude $\mathcal{E}({\bf r})$, from the polarization 
\begin{equation}
\mathcal{P}(\bf r)=\chi^{(1)}\mathcal{E}(\bf r)+\int {\rm d}{\bf r}^\prime \chi^{(3)}(|{\bf r}-{\bf r}^\prime|)|\mathcal{E}({\bf r}^\prime)|^2\mathcal{E}({\bf r}),
\end{equation}
by solving the many-body steady state of the driven interacting excitons to leading order in the Rydberg-state densities. As detailed in the supplementary material (section \ref{sec:nonlinear_optical_response}), this yields the linear response $\chi^{(1)}$ and permits to deduce the third order nonlinear susceptibility $\chi^{(3)}$ from the complex structure of the interaction potentials as shown in Fig.\ref{fig1}c-d. 
Its real part defines an effective photon-photon interaction $W(r)\approx \frac{\Omega^2}{4g} \chi^{(3)}_{R}(r)$, which can indeed be highly nonlocal (Fig.\ref{fig1}d) and extends over several hundred Bohr radii $a_b$ -- the characteristic length scale of interactions between ground state excitons \cite{ferrier2011}.
The characteristic soft core shape of $W(r)$ is consistent with the described blockade mechanism and features a strength reaching up to several tenths of $eV$, which in part arises from the high excitonic fraction, $\approx 4g^2/(4g^2+\Omega^2)$, of the dark-state polaritons formed under EIT conditions \cite{fleischhauer2005}. Remarkably, this behaviour persists in a regime where the above simple picture of a single blockaded Rydberg state breaks down entirely and the broadened Rydberg-exciton lines cover many interacting pair-states that are shifted to near-resonant energies (Fig.\ref{fig1}c). This can be traced back to a strong redistribution of the cavity coupling strength by the dipole-dipole interaction over a large number of Rydberg states that ultimately inhibits even the near resonant excitation of interacting Rydberg excitons. It is this "dilution blockade" that facilitates the emergence of strong photon interactions for broad Rydberg-excitation lines, as typically neglected for atomic systems but fully accounted for in the present formalism.

\begin{figure}
\includegraphics[width=0.99\linewidth]{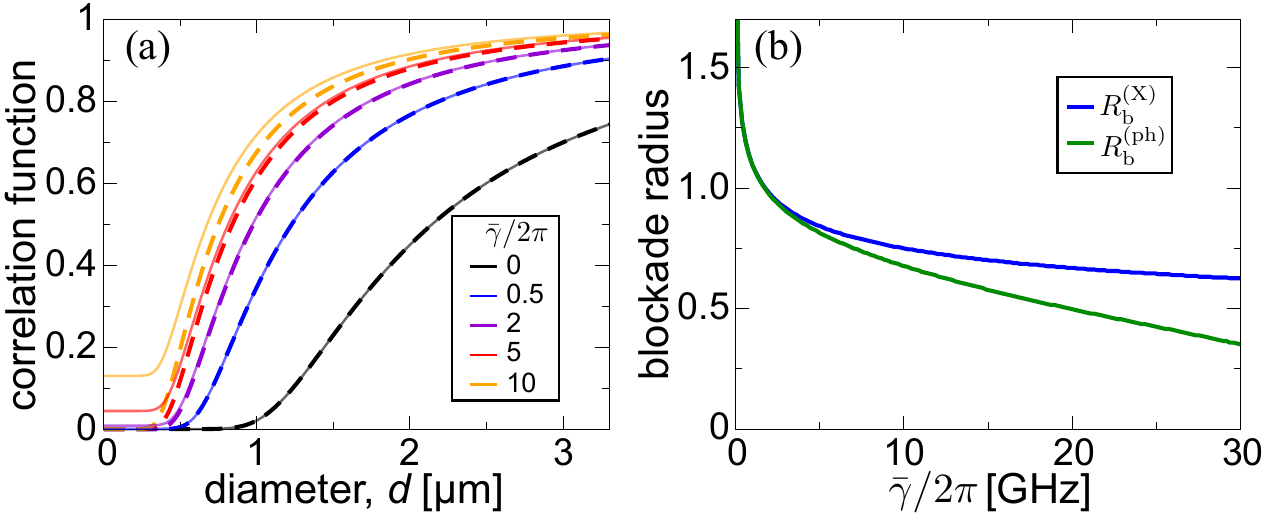}
\caption{(a) Equal-time photon-photon correlations $g^{(2)}(0)$ of the transmitted light as a function of the Rydberg-state linewidth, $\bar{\gamma}$, and the diameter, $d$, of the illumination area. (b) Critical photon-blockade radius $R_{\mathrm{b}}^{(\mathrm{ph})}$ for which $g^{(2)}(0)$ crosses $0.5$ (green) for different $\bar\gamma$ and fixed $\frac{\Omega^2}{|\Gamma|} = 40$GHz. The blue line shows the corresponding size $R_{\mathrm{b}}^{(\mathrm{X})}$ at which the probability to excite two Rydberg-excitons within the illumination area is suppressed by a factor of $2$. This exciton blockade radius scales as $\sim\bar{\gamma}^{-1/6}$.}
\label{fig3} 
\end{figure}

The nonlinearity can be probed experimentally by measuring the associated frequency shift $\delta_{\rm nl} \approx \frac{4 g^2}{\Omega^2}\alpha|\mathcal{E}|^2$ of the cavity transmission line. Here $\alpha=\int {\rm d}r^2 W(r)$ characterizes the effective photon interaction, in equivalence to the effective polariton scattering length arising from collisional exciton interactions, as observed, e.g., in GaAs \cite{ferrier2011}. As shown in Fig.\ref{fig2}a, the interaction strength achievable with Rydberg excitons exceeds that of collisional interactions by several orders of magnitude and strongly increases with the principal quantum number. 

While the finite Rydberg-state linewidth can often be neglected for atomic systems \cite{saffman2010}, it may present a major limiting factor for solid-state settings. However, as shown Fig.\ref{fig2}b, the vast enhancement of the optical nonlinearity even persists for considerable values of $\bar{\gamma}\approx0.1$meV, which is one order of magnitude higher than would be expected from pure radiative decay in WSe$_2$ monolayers \cite{moody2015}. Equally important, the enormous strength of the nonlinearity makes it possible to operate at such low probe-light intensities that additional exciton-density dependent effects \cite{moody2015,haugkoch2009} would not degrade the coherence of the system in the present situation.

Having established the emergence of a strong nonlinear response to weak coherent light fields in  spatially extended geometries, we can finally explore its effect on a few-photon quantum level. To this end, we now consider the opposite limit of a cavity that is coherently driven over a small illumination area with diameter $d$ (cf. supplementary materials, section \ref{sect:phot_correlations}). For sufficiently small $d$ the cavity can only accommodate a single Rydberg-exciton due to either or both of the blockade mechanisms described above. As a result, a single cavity-polariton will expose the high optical response of the low-lying exciton transition and ultimately block photon transmission for a sufficiently strong cavity coupling. As shown in Fig.\ref{fig3}a, the effective photon-photon interactions are indeed strong enough to alter the photon statistics over large distances and generate nonclassical states with strongly suppressed zero-time photon correlations, $g^{(2)}(0)<1$, for up to $d\sim1\mu$m. The collectively enhanced cavity-coupling to the spatially extended illumination area even enables single-photon transmission with $g^{(2)}(0)\approx0$, while increasing $\bar{\gamma}$ lowers Rydberg excitation and thereby gradually degrades the photon blockade induced by the single delocalized exciton. Importantly, however, the exciton blockade is largely unaffected by Rydberg state decay and decoherence, with a blockade radius that decreases only weakly as $R_{\rm b}^{({\rm X})}\sim \bar{\gamma}^{-1/6}$ and remains on a $\mu$m scale (Fig.\ref{fig3}b).
For Cu$_2$O \cite{kazimierczuk2014}, the stronger interactions and longer Rydberg-state lifetimes suggest even larger interaction ranges of $R_{\rm b}^{({\rm X})}\approx 3.1\mu$m for $n=10$ and $R_{\rm b}^{({\rm X})}\approx 14.6\mu$m for $n=20$ in an equivalent setup.

Our results show that Wannier Rydberg excitons provide an intrinsic mesoscopic lengthscale capable of supporting collective excitations that permit the all-optical manipulation of light. 
Yet, this photon interaction-range can be much smaller than typical system sizes, which opens up a new regime of photonic many-body physics, well beyond the capabilities of corresponding atomic systems. Here, the strong achievable light-matter coupling, the demonstrated valley-dependent interactions and the special Bloch-band geometry of TMDC materials provide perspectives for exploring multi-component systems and topological states with many strongly interacting photons. 
On the other hand, the exaggerated properties of Rydberg-excitons \cite{heckotter2017} combined with the integrability and continually advancing functionalities of low-dimensional semiconductors \cite{tonndorf2015,He2016} hold promise for few-photon \cite{chang2014} applications.
Already on a classical level, the typical time and energy scales of the described system may enable fast optical switching at ultralow light-intensities, and the exploration of new collective nonlinear phenomena in exciton-polariton condensates.

This work was funded by the EU through the H2020-FETPROACT-2014 grant number 640378 (RYSQ), by the DFG through the SPP 1929 and by the DNRF through a Niels Bohr Professorship to T.P..

\onecolumngrid
\appendix

\section{Single Exciton Solution} \label{sect:single_exciton}
Semiconductor excitons are bound electron hole pairs whose binding energy reduces the free electron band gap to the quasi-particle band gap, which is important and in some cases dominant for the optical properties.
Both electron and hole have intricate dispersion relations inherent to the material's band structure.
Transition metal dichalcogenide (TMDCs) are direct semiconductors in the monolayer limit with least energy transitions 
at the $K$ points, the corners of the hexagonal Brillouin zone \cite{splendiani2010emerging}.
Though monolayer TMDCs resemble graphene in many ways, their band structures differ in (at least) one mayor feature:
While graphene exhibits Dirac cones at $\pm K$, the valence and conduction bands in TMDCs are split naturally by a sizable band gap, 
rendering the dispersions quadratic for small reciprocal vectors $\vec{k}$ off the $K$ points.
Excitons in TMDCs are formed at the $\pm K$ points and have a large binding energy on the order of $0.5$meV, making them stable against thermal fluctuations even at room temperature.
A full theoretical description requires ab-initio methods, but an accurate and very insightful description is given by the standard effective model  \cite{srivastava2015, zhou2015}
\begin{align}
  2 E_{\vec{k}} f(\vec{k}) + \sum_{\vec{k}^{\prime}} f(\vec{k}^{\prime}) V_{\vec{k},\vec{k}^{\prime}} \Braket{c\vec{k}| c\vec{k}^{\prime}} \Braket{v\vec{k}|v\vec{k}^{\prime}} = E f(\vec{k}),
\end{align}
where $E_{\vec{k}}$ is the electron/hole dispersion, $f(\vec{k})$ are the amplitudes of creating a exciton at relative wavevector $\vec{k}$ 
and $V_{\vec{k}, \vec{k}^{\prime}}$ are the Fourier components of the electron-hole potential $V_{eh}(r)$. 
The Bloch states overlaps $\Braket{c\vec{k}| c\vec{k}^{\prime}}$ (conduction band) and $\Braket{v\vec{k}|v\vec{k}^{\prime}}$ (valence band) imprint the crystal topology onto the excitons. 
In an expansion around $\pm K$ these overlaps can be expressed as $\vec{k}$ dependent functions of the Berry curvature $\Omega_{0}$ with the real space analogs \cite{zhou2015}
\begin{align} \label{eq:berry_1}
 V = V_{eh}(r) + \frac{- \tau |\Omega_{0}|}{2 \hbar} (\nabla V_{eh} \times \vec{p})_{z} + \frac{|\Omega_{0}|}{4} \nabla^{2}V_{eh}(r)
\end{align}
where $\tau = \pm 1$ denotes the valley index and $\vec{p}$ is the momentum operator.
Monolayer TMDCs being virtually two-dimensional (2d) materials, the interesting situation arises in which the excitonic wavefunction is confined to the plane, while 
electromagnetic interactions can also enter the surrounding environment, i.e. are three-dimensional (3d).
As a result, the potential is screened at short separations between electron and hole (as is typical in bulk materials), but essentially
unscreened at larger distances. Following the model in \cite{cudazzo2011dielectric, andryushin1980electron, chernikov2014, keldysh1979jetp, rytova1967screened, chaplik1971absorption} of this situation we use the well-established potential
\begin{align} \label{eq:v_eff}
  V_{eh}(r) = -\frac{e^2}{4\pi \epsilon_{0}} \frac{\pi}{2r_{0}} \left[ H_{0}(r/r_{0}) - Y_{0}(r/r_{0})\right], 
\end{align}
where $H_{0}$ is the first Struve function and $Y_{0}$ is first Bessel function of the second kind and $r_0$ is the effective screening length (taken from \cite{chernikov2014}),
capturing the cross-over between Coulomb behavior at large $r$ and a (weaker) logarithmic decay at small $r$.
Introducing center-of-mass and relative coordinates
\begin{equation}
\begin{aligned} \label{eq:rel_coord}
 \vec{R}_{i} &= \frac{m_{e}\vec{r}_{ei}+m_{h}\vec{r}_{hi}}{m_{e} + m_{h}}, \qquad \vec{r}_{i} &= \vec{r}_{ei} - \vec{r}_{hi}.
\end{aligned}
\end{equation}
the Schr\"odinger equation for the relative coordinate reads
\begin{align}
 -\frac{\hbar^2}{2 \mu}  \Delta \psi - \left( E - V(r) \right) \psi = 0,
\end{align}
where $\mu = \frac{m_{e}m_{h}}{m_{e}+m_{h}}$ is the reduced mass. A product ansatz $\psi(\vec{r}) = \rho(r) \Phi(\phi)$ gives
\begin{align}
 -\frac{\hbar^2}{2 \mu} \left[ \frac{r^2}{\rho} \frac{\partial^2 \rho}{\partial r^2} + \frac{r}{\rho} \frac{\partial \rho}{\partial r} + \frac{1}{\Phi} \frac{\partial^2 \Phi}{\partial \phi^2}\right] - (E - V(r))r^{2} = 0,
\end{align}
where we refer to $m$ as the ``angular'' quantum number (cf. below for comparison with 3d case),
leading directly to the orbital eigenfunctions
\begin{align}
 \Phi(\phi) = \frac{1}{\sqrt{2 \pi}} e^{i m \phi}, \ m \in \mathbb{Z}.
\end{align}
The radial equation then reads
\begin{align} \label{eq:hydro_radial}
 \frac{d^2 \rho}{dr^2} + \frac{1}{r} \frac{d\rho}{dr} + \left[ \frac{2\mu}{\hbar^2} \left( E - V(r) \right) - \frac{m^2}{r^2} \right]\rho = 0.
\end{align}
We cast the radial solutions as $u(r) = r^{\beta} \rho(r)$, transforming the radial equation into
\begin{align}
 \frac{d^2 u}{d r^2} = \left[ -\frac{2 \mu}{\hbar^2} (E-V(r)) + \frac{m^2-\beta^2}{r^2} \right]u(r) + \frac{2\beta-1}{r} \frac{d u(r)}{d r}.
\end{align}
Standard numerical methods work fine for $m>0$ when the term $(m^2-\beta^2)/r^2$ constitutes an effective repulsive potential.
However, for $m = 0$ this term becomes attractive and at small distances it dominates the Coulomb potential $-\beta^2/r^2 \ll -1/r \ \text{if} \ 0 < r < \delta$.
The resulting equation is that of a one-dimensional Schrödinger equation with a potential $V(\rho) \propto 1/\rho^2$.
This potential has very unusual properties \cite{essin2006quantum} because it is just on the boundary of permitting bound states (those are impossible for stronger potentials $V \propto 1/r^{2+\epsilon}$).
We follow a recently proposed algorithm \cite{pikovski2014differentiation} which suggests an Euler integrator anticipating the correct solution. 

\begin{figure}
\begin{center}
 \includegraphics[height=0.99\textwidth]{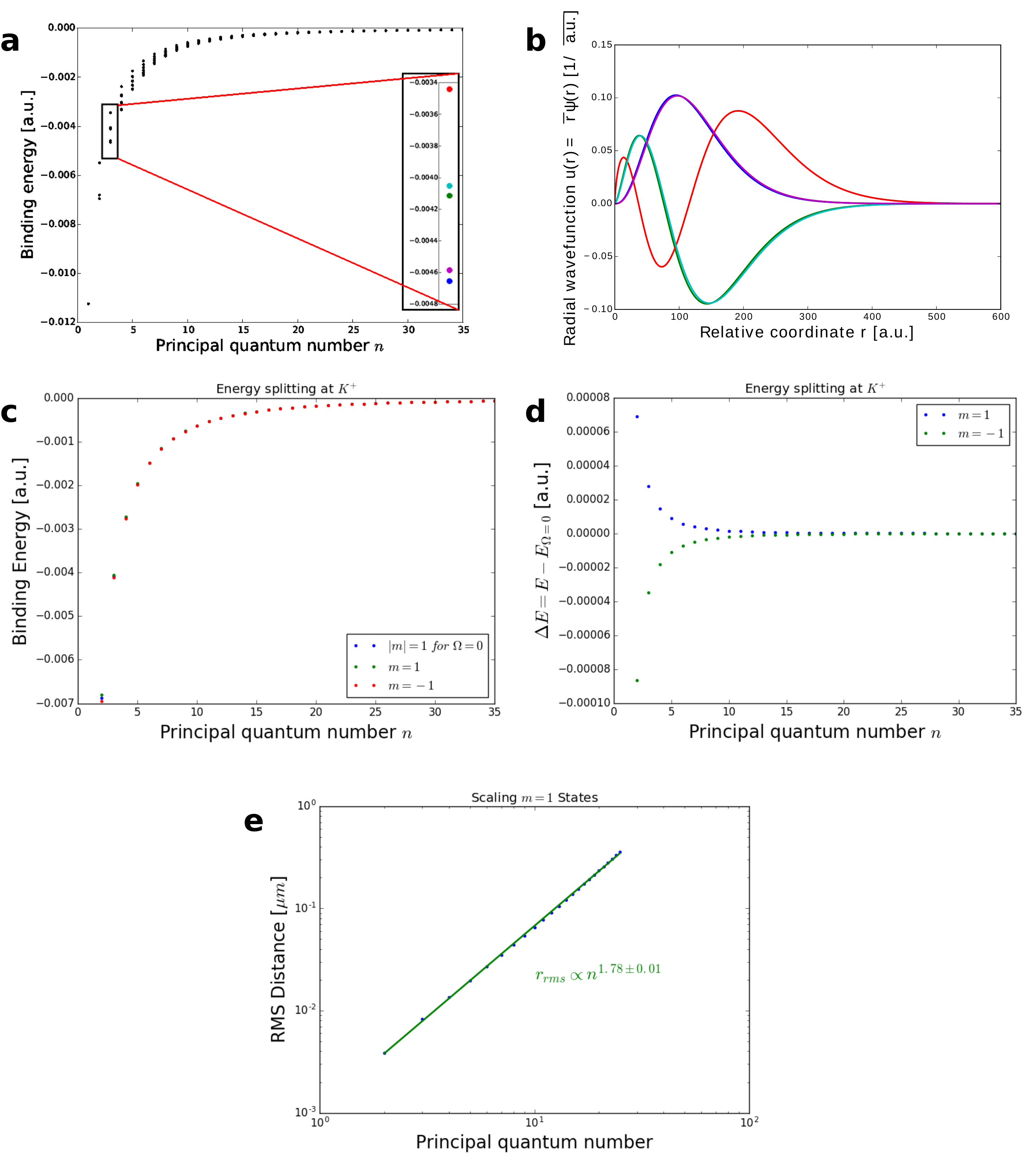}
\end{center}
\caption{ a) Excitonic spectrum at $K$, energies and wavefunctions singled out for $n=3$ (colors matching). Negative $m$ are slightly lower in energy than positive $m$.
b) Wavefunctions corresponding to a) (colors matching), c/d) Energy splittings due to the Berry terms, corrections are small. e) The root mean square electron-hole separation $r_{\text{rms}}$ scales less strongly than $n^{2}$. \label{fig:exciton_spectrum}}
\end{figure}

Let us now comment on the physical implications of this exciton model:
The excitonic wavefunction is fully characterized by the the principal quantum number $n$ determined from the radial solution and and the angular part $m$.
Compared to the 3d Coulomb case, the quantum number $l$ is ``frozen'' at its maximum value, thus leaving only $n$ and $m$ \cite{zaslow1967two}.
The latter determines the optical properties, such that we refer to $m=0$ as s-states, $|m|=1$ as p-states etc.
While the exact $1/r$ Coulomb potential leads to a complete degeneracy of all states at given $n$, the screened potential $V_{eh}$ splits the energy states according to $|m|$.
Contrary to expectations based on the 3d counterpart, this (quantum) defect scales almost linearly with $|m|$,
allowing to optically address states other than the s-states (Fig.~S\ref{fig:exciton_spectrum}).
The energy shifts are quite pronounced and dominate the non-hydrogenic nature of the spectrum \cite{chernikov2014}, whereas the second and third terms in Eq. (\ref{eq:berry_1}) are small corrections. 
They are topological terms, accounting for a combination of strong spin-orbit coupling in TMDCs and the breaking of inversion symmetry which results in the six $K$-points falling into two different classes $\pm K$ \cite{xiao2012coupled},
characterized by Berry curvatures $\pm \Omega_{0}$. While both Berry terms are rotationally symmetric and, thus, 
do not mix (bare) states of different $m$, the first term acts like a position dependent magnetic field $(\nabla V \times \vec{p})_{z} = B(r)L_{z}, B(r)=|\nabla V|$, lifting the degeneracy in $\pm m$.
Note that this breaking of time reversal symmetry is opposite at $\pm K$, restoring overall time reversal symmetry. 
This simplified expansion around $K$ was first proposed for $n=2$ where the energy splitting is largest \cite{srivastava2015}. 
We emphasize that the approximation improves as higher quantum numbers are considered.

TMDCs feature valley spin-orbit coupling of the valence and (to a lesser extent) the conduction band, such that the bands acquire different energy shifts and are susceptible 
to either $\sigma^{+}$ or $\sigma^{-}$ light \cite{xiao2012coupled}. By using either of the circularly polarized types of light, we can restrict our attention
to the quadratic regions close the $+K$ or $-K$ points in the Brillouin zone.
For the intra-excitonic transitions to the Rydberg states optical selection rules arise directly from the excitonic wavefunctions:
only transition with $m^{\prime} = m \pm 1$ are dipole-allowed.
In particular, we can select $m^{\prime} = m+1$ using $\sigma^{+}$ light and $m^{\prime} = m-1$ by using $\sigma^{-}$ light.

\section{Excitonic Pair Interactions} \label{sect:dipole_interaction}
Excitonic ground state interactions are typically of exchange character: The wavefunctions of two excitons overlap
giving rise to interactions based on the fermionic nature of the electrons. 
If addressing Rydberg states, excitons are usually separated by hundreds of Bohr radii, such that the primary type of interaction
is electromagnetic. At such great distances $r$ screening effects become negligible for the aforementioned reasons
\begin{align}
 \lim_{r/r_{0} \gg 1} V_{eh}(r) \rightarrow -\frac{e^{2}}{4 \pi \epsilon_{0}} \frac{1}{r} \left[ 1 + \mathcal{O}((r/r_{0})^{-2}) \right]
\end{align}
and the resulting Coulombic terms 
\begin{align} \label{eq:dipole_operator}
 V^{dd} = \frac{e^{2}}{4 \pi \epsilon_{0}} \left(- \frac{1}{r_{h_{1}e_{2}}} - \frac{1}{r_{e_{1}h_{2}}} + \frac{1}{r_{h_{1}h_{2}}} + \frac{1}{r_{e_{1}e_{2}}} \right),
\end{align}
where $r_{h_{1}e_{2}}$ is the separation of the first exciton's electron from the second exciton's hole and so on,
are expanded into a standard van der Waals potential with respect to the center of mass coordinates, using expansions of the form
\begin{align} \label{eq:dipole_intermediate}
 \frac{1}{r_{h_{1}e_{2}}} \approx \frac{1}{R} \left( 1 - \frac{1}{MR} q^{z}_{e_{1}h_{2}} -\frac{1}{2} \left[\frac{1}{RM} \vec{q}_{e_{1}h_{2}} \right]^2 + \frac{3}{8}\left[ \frac{2}{MR} q^{z}_{e_{1}h_{2}} \right]^2 \right),
\end{align} 
where $\vec{R} = \vec{R}_{1} - \vec{R}_{2}$ is the vector connecting the excitons' centers of mass, $\vec{q}_{e_{1}h_{2}} = m_{h}\vec{r}_{1} + m_{e}\vec{r}_{2}$ and $q^{z}_{e_{1}h_{2}} = \vec{q}_{e_{1}h_{2}} \vec{R}/R$ is the projection of this vector on $\vec{R}$. We arrive at the well-known dipole-dipole interaction (in relative and center of mass coordinates) with $z_{i}=\vec{r}_{i}\vec{R}/R$
\begin{equation}
\begin{aligned}
 V^{dd} &= \frac{e^{2}}{4 \pi \epsilon_{0}} \frac{1}{M^2 R^3} \left[ M^2 \vec{r}_{1} \vec{r}_{2} - 3M^2 z_{1} z_{2} \right] = \frac{e^{2}}{4 \pi \epsilon_{0}} \frac{1}{R^3} \left[\vec{r}_{1} \vec{r}_{2} - 3 z_{1} z_{2} \right].
\end{aligned}
\end{equation}
The excitonic pair Hamiltonian
\begin{align} \label{eq:ex_hamiltonian_plus_interaction}
 \hat{H} = \hat{H}^{ex}_{1} + \hat{H}^{ex}_{2} + \hat{V}^{dd}
\end{align}
acts on the Hilbert space spanned by the product states $\{ \Ket{\alpha} = \Ket{\psi_{n_1 m_1}; \psi_{n_2 m_2}} \equiv \Ket{\psi_{i}, \psi_{j}} \}$. 
This basis diagonalizes the single-particle contributions for the first exciton
\begin{align}
 \hat{H}^{ex}_{1} = \sum_{n, m} E_{n,m} \Ket{\psi_{n,m}(\vec{r}_{1})}\Bra{\psi_{n,m}(\vec{r}_{1})} \otimes \mathbb{1},
\end{align}
where $E_{n,m}$ is the energy of state $\Ket{n,m}$, and likewise for the second exciton but it is coupled by the off-diagonal terms in 
\begin{equation}
\label{eq:dipole_operator_matrix}
\begin{aligned}
 V^{dd} &= \frac{V^{dd, rad}}{4} \left[ -(\delta_{m_{1},m_{1}^{\prime}+1}\delta_{m_{2}+1,m_{2}^{\prime}} + \delta_{m_{1}+1,m_{1}^{\prime}}\delta_{m_{2},m_{2}^{\prime}+1}) \right.\\
  -3( &\left. e^{-2i\theta} \delta_{m_{1}, m_{1}^{\prime}+1} \delta_{m_{2}, m_{2}^{\prime}+1} + e^{2i\theta} \delta_{m_{1}+1, m_{1}^{\prime}} \delta_{m_{2}+1, m_{2}^{\prime}})\right], \\
  V^{dd, rad} &= \Braket{\rho_{n_{1}m_{1}}(\vec{r}_{1})|\vec{r}_{1}|\rho_{n_{1}^{\prime}m_{1}^{\prime}}(\vec{r}_{1})} \cdot \Braket{\rho_{n_{2}m_{2}}(\vec{r}_{2})|\vec{r}_{2}|\rho_{n_{2}^{\prime}m_{2}^{\prime}}(\vec{r}_{2})},
\end{aligned}
\end{equation}
where we can w.l.o.g. choose the absolute in-plane orientation $\theta = 0$, $V^{dd, rad}$ falls off with $R^{-3}$ and contains the dipole matrix elements. 
The numerical diagonalization is facilitated by the symmetries of the excitonic Hamiltonian. 
First, as is obvious from the pair basis representation, $V^{dd}$ does not couple states with even and odd $M = m_{1} + m_{2}$.
Second, the entire Hamiltonian is invariant w.r.t. particle exchange, implying that the Hilbert space can be decomposed into 
non-interacting subspaces of odd and even states. 
In a LCAO-type numerical procedure, we diagonalize the excitonic spectrum in the subspace of symmetric states with even $M$ with a LAPACK algorithm for sparse matrices, 
ensuring convergence by varying the basis set. 

By diagonalizing the Hamiltonian from Eq. (\ref{eq:ex_hamiltonian_plus_interaction}) we obtain the excitonic interaction potential surfaces $U_{\mu}$, as shown for example in Fig. 1c in the main text.
As the van der Waals interaction increase enormously with $n$, it couples only excited states and leaves the excitonic ground states unchanged.
Thus, we summarize the diagonalization formally in the unitary transformation
\begin{align} \label{eq:diagonalizing_tranformation}
 \Ket{\mu} = \hat{U}^{\dagger} \Ket{\alpha} = \sum_{\alpha} c_{\alpha, \mu}^{*} \Ket{\alpha},
\end{align}
restricting $\Ket{\alpha}$ to the manifold of doubly excited states.

\section{Total Hamiltonian and Equations of Motion} \label{sect:hamiltonian}
We consider an infinite system of excitons in a cavity. Each exciton has a vacuum state, an excitonic ground state $g$ 
(corresponding to $\Ket{1s}$ and created by $\hat{X}^{\dagger}_{g}$) at $\hbar \omega_{g}^{ex}$ and a number of 
interacting Rydberg states $\{ k \}$ (corresponding to $\Ket{np}$ and created by $\hat{X}^{\dagger}_{k}$) at energies $\hbar \omega_{k}^{ex}$. The excitons are driven by a weak cavity field (operator $\hat{\mathcal{E}}$) at frequency $\omega_{p}$ and a strong external laser field.
We consider a single longitudinal cavity mode (for simplicity assume the lowest energy mode), such that we can describe the in-plane light propagation via a Schrödinger equation with effective mass $m_{ph}=\hbar nk_{z}/c$ \cite{carusotto2013},
where $k_{z} = \frac{\pi}{l_{z}}$, $l_{z}$ is the cavity width, $n$ is the dielectric constant in the cavity and $c$ is the vacuum speed of light.
This picture is valid if $k/k_{z} \ll 1$ (quadratic band) and $\frac{\hbar^{2}k^{2}}{2m} \ll \frac{2 \pi \hbar c}{nl_{z}} \leftrightarrow k \ll \frac{2 \pi}{l_{z}}$ (large separation of odd modes), 
which is (practically) equivalent to the first condition.
As is standard in the low-excitation regime (discussed in detail below), we bosonize the spin operators, such that the Hamiltonian for the case of a coherent driving field $\hat{E}^{\text{in}}(\vec{r},t) = \hat{E}^{\text{in}}_{0} e^{-i \omega_{\text{in}}t}$ with coupling constant $\eta$
reads, in rotating wave approximation $\hat{H} = \int d^2\vec{r} \hat{h}(\vec{r})$ with
\begin{equation}
 \begin{aligned}
 \hat{h}(\vec{r}) &= \eps \left(- \frac{\hbar^{2}}{2 m_{ph}} \nabla^{2} + \hbar \omega_{cav} \right)\epsdag  \\
 &+ \gsdag \left( - \frac{\hbar^{2}}{2 m_{ex}} \nabla^{2} + \hbar \omega_{g}^{ex} \right)\gs \\
 &+ \sum_{i} \grydag{i}(\vec{r}) \left( - \frac{\hbar^{2}}{2 m_{ex}} \nabla^{2} + \hbar \omega_{i}^{ex} \right) \gry{i}(\vec{r}) \\
 &+ \hbar g \left( \epsdag \gs + h.c. \right) - \sum_{i} \hbar \frac{\Omega_{i}}{2} \left( \hat{X}_{g}^{\dagger}(\vec{r})\hat{X}_{i}(\vec{r}) + h.c. \right) \\
 &+ \sum_{i \leq j, i^{\prime} \leq j^{\prime}} \int d\vec{r}^{\prime} \grydag{i^{\prime}}(\vec{r}) \grydag{j^{\prime}}(\vec{r}^{\prime}) V^{dd}_{i^{\prime}j^{\prime},ij} (|\vec{r} -\vec{r}^{\prime}|) \gry{i}(\vec{r}) \gry{j}(\vec{r}^{\prime})\\
 &+ i\epsdag \eta \hat{E}(\vec{r}) + h.c.,
 \end{aligned}
\end{equation}
where $\eps$ and $\gs$ rotate at $e^{i\omega_{\text{in}}t}$ with respect to the lab frame, 
while $\gry{k}$ rotates at $e^{i (\omega_{\text{in}}+\omega_{c})t}$ relative to the lab frame.
Summarizing the previous sections, the first, second and third terms represent the free in-plane motion of the cavity field, the ground state and excited state excitons, respectively.
The fourth line contains the laser coupling between vacuum and ground state excitons as well as the subsequent excitation to Rydberg levels (at driving frequency $\omega_{c}$),
while the second to last line describes dipole coupling between pairs of excitons, inducing many-body correlations.
Balancing the pumping (final line), the excitonic dynamics is naturally subject to decoherence processes: Excited states may spontaneously decay to lower and dipole coupled states, 
but there are also dephasing mechanisms attenuating quantum coherences. Assuming $\delta$-correlated decoherence mechanisms, 
we model both types of processes by single-particle Lindblad master equations $\mathcal{L}(\rho) = \hat{L}^{\dagger}\rho\hat{L} + \frac{1}{2}\left( \hat{L}^{\dagger}\hat{L}\rho + \rho\hat{L}^{\dagger}\hat{L} \right)$.
In the limit of very small excitation fractions decay of excitons can be modeled as dephasing. Additionally, photons may leak out of the 
semiconductor cavity through imperfect mirrors at a loss rate $\kappa$.
Within the used approximations, we can use an effective Hamiltonian to formulate Heisenberg equations of motion
\begin{equation}
\begin{aligned}
 \partial_{t} \eps &= i\omega_{\text{in}} \eps -i\left( \omega_{\text{cav}} - \frac{\hbar}{2 m_{ph}} \nabla^{2} \right) \eps - i g \gs 
 - \frac{\kappa}{2} \eps + \eta \hat{E}^{\text{in}}(\vec{r}) \\
\end{aligned}
\end{equation}
\begin{equation}
\begin{aligned}
 \partial_{t} \gs &= i \omega_{\text{in}}\gs -i g \eps - i \left( - \frac{\hbar}{2 m_{ex}} \nabla^{2} + \omega_{g}^{ex} \right)\gs 
 + i \sum_{i} \frac{\Omega_{i}}{2} \gry{i}(\vec{r}) - \frac{\gamma}{2} \gs \\
\end{aligned}
\end{equation}
\begin{equation}
\begin{aligned}
\partial_{t} \gry{k}(\vec{r}) &= i (\omega_{\text{in}} + \omega_{c}) \gry{k}(\vec{r}) + i \frac{\Omega_{k}}{2} \gs -i \left( - \frac{\hbar}{2 m_{ex}} \nabla^{2} + \omega_{k}^{ex} \right) \gry{k}(\vec{r}) \\
&- i\sum_{i \leq j, i^{\prime}} \int d^{2}r^{\prime} \grydag{i^{\prime}}(\vec{r}^{\prime}) \frac{V^{dd}_{i^{\prime}k,ij}(|\vec{r}^{\prime} - \vec{r}|)}{\hbar} \gry{i}(\vec{r}^{\prime}) \gry{j}(\vec{r}) - \frac{\bar{\gamma}_{k}}{2} \gry{k}(\vec{r}).
\end{aligned}
\end{equation}
We introduce the following notation
\begin{equation}
\begin{aligned}
 \Delta_{\text{cav}} &\equiv \omega_{\text{in}} - \omega_{\text{cav}} \qquad \Gamma_{\text{cav}} \equiv \kappa -i2\Delta_{\text{cav}} \\
 \Delta &\equiv \omega_{\text{in}} - \omega_{g}^{ex} \qquad \Gamma \equiv \gamma - i2\Delta \\
 \Delta_{k} &\equiv \omega_{\text{in}} + \omega_{c} - \omega_{k}^{ex} \qquad \Gamma_{k} \equiv \bar{\gamma}_{k} - i2\Delta_{k}.
\end{aligned}
\end{equation}
and eliminate the intermediate state
\begin{align} \label{eq:adiabatic_elim_basic}
 \gs = 2i \frac{-2 g \eps  +  \sum_{i} \Omega_{i} \gry{i}(\vec{r})}{2 \Gamma}.
\end{align}
We remark that the adiabatic elimination of Eq. (\ref{eq:adiabatic_elim_basic}) is exact (in the steady state) only for non-interacting excitons.
The approximation we make by taking it over to the interacting many-body system works well if $\Omega \ll \sqrt{\Delta^{2} + \gamma^{2}}$ \cite{PhysRevLett.116.243001}.
Note that this is last step reduces complexity from the equations but it is no fundamental limitation as the intermediate state can be treated fully.
The final equations of motion read
\begin{align} \label{eq:light_field}
  \partial_{t} \eps &= -\frac{\Gamma_{\text{cav}}}{2}\eps + i\frac{\hbar}{2 m_{ph}} \nabla^{2} \eps - \frac{2g^{2}}{\Gamma}\eps + \frac{g}{\Gamma} \sum_{i} \Omega_{i} \gry{i}(\vec{r})  + \eta \hat{E}^{\text{in}}(\vec{r})
\end{align}
\begin{equation} \label{eq:adiabatic_elim}
 \begin{aligned}
	 \partial_{t} \gry{k}(\vec{r}) &= \frac{g\Omega_{k}}{\Gamma} \eps - \frac{\Gamma_{k}}{2} \gry{k}(\vec{r})
  -\frac{\Omega_{k}}{2\Gamma} \sum_{i} \Omega_{i} \gry{i}(\vec{r}) 
  - i\sum_{i \leq j, i^{\prime}} \int d\vec{r}^{\prime} \grydag{i^{\prime}}(\vec{r}^{\prime}) \frac{V^{dd}_{i^{\prime}k, ij}(|\vec{r}^{\prime} - \vec{r}|)}{\hbar} \gry{i}(\vec{r}^{\prime}) \gry{j}(\vec{r}).
\end{aligned}
\end{equation}
These equations fully specify the temporal evolution of the quantum many-body system and are the starting point of all further calculations.

\section{Nonlinear Optical Response} \label{sec:nonlinear_optical_response}

In the following section, we derive an effective nonlinear propagation equation for the in-plane cavity field.
As can be seen from Eqs. (\ref{eq:light_field}-\ref{eq:adiabatic_elim}), this requires, at least in principle, to obtain the full evolution of $\hat{X}_{k}$,
which couples to the entire many-body problem of excitons and spins.
Instead, we restrict our attention to a set of physical conditions which affords three approximations permitting us to find a closed analytic solution.
First, we consider the case of coherent light, allowing us to neglect field-spin correlations $\mathcal{\hat{E}}\hat{X} \rightarrow \mathcal{E} \hat{X}$.
Second, we exploit that the internal excitonic dynamics evolve on a time scale much faster than the in-plane light propagation, 
such that we treat $\mathcal{E}(\vec{r})$ as an adiabatic parameter in solving the many-body excitonic steady state.
Third, we focus on the limit of very small Rydberg excitation fractions, which can be achieved by weak cavity fields and is later checked self-consistently.
This assumption implies that excitonic motion (i.e. the external degree of freedom) driven by dipole forces can be neglected and it is justified to solve the dynamics for a uniform distribution of excitons. 
Note that the internal degrees of freedom, the excitonic polarization, is position dependent and drives the optical response.
We start our considerations by expanding the hierarchy of excitonic equations of motion, 
tracing the dependence of the main correlator $\grydag{i^{\prime}}(\vec{r}^{\prime}) \gry{i}(\vec{r}^{\prime}) \gry{j}(\vec{r})$ from Eq. (\ref{eq:adiabatic_elim}) to terms of the form $\gry{l}(\vec{r}^{\prime})\gry{k}(\vec{r})$, $\gry{l}^{\dagger}(\vec{r}^{\prime}) \gry{k}(\vec{r})$
as well as the single-particle element $\gry{k}^{\dagger}(\vec{r})\gry{k}(\vec{r})$.
Besides single- and two-particle terms, couplings to three-particle correlators enter, which couple to four-particles correlators and so on.
However, we are interested in the limit of weak driving fields, where only very few excitons are excited to their Rydberg state (cf. third step above). 
As the density of Rydberg states decreases,
higher-order correlations become less significant and at sufficiently low densities we are justified in truncating the series at pair correlators.
The regime of low densities is reached as soon as the expectation value of Rydberg excitons per photonic blockaded sphere is small.
After taking expectation values and considering the excitonic adiabatic steady state (cf. second step above), we are left with a closed set of algebraic equations.
The key step is to transform the steady state equations from the product basis to the pair basis (Eq. (\ref{eq:diagonalizing_tranformation})), as for example
\begin{equation}
\begin{aligned}
  &\sum_{l, k^{\prime}, l^{\dprime}} V^{dd}_{kl, k^{\dprime} l^{\dprime}} \langle \gry{k^{\prime}}(\vec{r}) (\gry{l}^{\dagger}\gry{l^{\prime}})(\vec{r}^{\prime}) \rangle 
  = \sum_{l} \sum_{\mu} \left[ U_{\mu}(|\vec{r}-\vec{r}^{\prime}|)  + (\Delta_{k} + \Delta_{l})  \right] c^{*}_{[kl], \mu} Y_{gl, \mu} (\vec{r}, \vec{r}^{\prime}),
\end{aligned}
\end{equation}
where $Y_{gl, \mu} (\vec{r}, \vec{r}^{\prime})$ is the expecation value of the operator destroying a pair state $\mu$ shared between excitons in $\vec{r}$ and $\vec{r}^{\prime}$ and
replacing it by a product state of a ground state and a Rydberg state $l$ in the same positions.
The insights gained from the new basis are twofold: Firstly, the equations become diagonal (the residual off-diagonal couplings are very weak) and can, thus, be solved analytically.
Secondly, the pair basis provides the correct energy scale to consider. Rather than using the unfeasible number of all states
we restrict our attention to states within an energy window around the two-photon resonance.
We eliminate two-particle terms and after some algebra $Y_{gl, \mu} (\vec{r}, \vec{r}^{\prime})$ is expressed only in terms of single-particle correlators.
Finally, we evaluate this explicit expression of the nonlinear response using the ansatz of the (approximate and non-interacting) EIT ground state 
$\langle \hat{X}_{k} \rangle = (2g \mathcal{E}(r)\Omega_{k})/(\Gamma\Gamma_{k} + \Omega_{k}^{2})$
and are now able to formulate a nonlinear equation for the medium's polarization $\mathcal{P}(\vec{r}) = \Braket{\gs}$ (Eq. (1) in main text)
\begin{align}
 \mathcal{P}(\vec{r}) = \chi^{(1)} \mathcal{E}(\vec{r}) + \int d\vec{r}^{\prime} \chi^{(3)}(|\vec{r} - \vec{r}^{\prime}|)|\mathcal{E}(\vec{r}^{\prime})|^2 \mathcal{E}(\vec{r})
\end{align}
The linear susceptibility $\chi^{(1)}$ is given by
\begin{align}
 \chi^{(1)} = -2g\frac{i}{\Gamma} \left[ 1 - \sum_{k}\frac{\Omega_{k}^{2}}{\Omega_{k}^{2} + \Gamma_{k}\Gamma} F_{k}^{(1)} \right] \qquad F_{k}^{(1)} = 1 - \sum_{k^{\prime} \neq k} \frac{\Omega_{k^{\prime}}^{2}}{\Omega_{k^{\prime}}^{2} + \Gamma_{k^{\prime}}\Gamma},
\end{align}
while the nonlinear susceptibility $\chi^{(3)}$ reads
\begin{align}
 \chi^{(3)}(|\vec{r}-\vec{r}^{\prime}|) &= -16 g^3\sum_{k, l} \frac{\Omega_{k} F_{k}^{(3)}}{\Omega_{k}^{2} + \Gamma_{k}\Gamma} \sum_{\mu} \left[ U_{\mu}(|\vec{r}-\vec{r}^{\prime}|)  + (\Delta_{k} + \Delta_{l})  \right] c^{*}_{[kl], \mu} Y_{gl, \mu} (|\vec{r} - \vec{r}^{\prime}|) \label{eq:chi3} \\\
 Y_{gl, \mu} (|\vec{r} - \vec{r}^{\prime}|) &= -\frac{\Omega_{l}}{\Omega_{l}^{2} + \Gamma_{l}^{*}\Gamma^{*}}  \cdot  \frac{ \sum_{k^{\prime} l^{\prime}} c_{[k^{\prime}l^{\prime}], \mu} \cdot A_{k^{\prime},l^{\prime}}}{\tilde{\Omega}^{2}(\mu) + \Gamma \left[ iU_{\mu}(|\vec{r}-\vec{r}^{\prime}|) + \sum_{k^{\prime} l^{\prime}} |c_{[k^{\prime} l^{\prime}], \mu}|^{2} \gamma_{k^{\prime}} \right]} \label{eq:coupling_1}
\end{align}
where we defined
\begin{align} \label{eq:aux_Y}
 F_{k}^{(3)} &= 1 - \Omega_{k} \sum_{k^{\prime} \neq k} \frac{\Omega_{k^{\prime}}}{\Omega_{k^{\prime}}^{2} + \Gamma_{k^{\prime}}\Gamma} \\
 A_{k^{\prime}, l^{\prime}}&= \frac{\Omega_{k^{\prime}} \Omega_{l^{\prime}}}{2} \cdot \frac{[\Omega_{k^{\prime}}^{2} + \Gamma_{k^{\prime}} \Gamma] + [\Omega_{l^{\prime}}^{2} + \Gamma_{l^{\prime}} \Gamma]}{[\Omega_{k^{\prime}}^{2} + \Gamma_{k^{\prime}} \Gamma] \cdot [\Omega_{l^{\prime}}^{2} + \Gamma_{l^{\prime}} \Gamma]} \\
 \tilde{\Omega}^{2}(\mu) &= \sum_{k, l, m} c_{[k l], \mu} c^{*}_{[k m], \mu} \Omega_{l} \Omega_{m}.
\end{align}
Both single-particle and two-particle terms have very weak off-diagonal coupling terms, which we capture perturbatively with the help of $F_{k}^{(1)}$ and $F_{k}^{(3)}$, both of which are typically quite small.
It turns out that there are only odd terms affecting the light field.
In fact, all even orders vanish as can be shown from the inversion symmetry of a medium. 
While one could object that TMDC materials break inversion symmetry leading to important features including second harmonic generation \cite{PhysRevLett.114.097403},
our results are restricted to the limit of Rydberg excitons, which span hundreds of thousands of crystal cells, and are, thus, very well approximated by (inversion-symmetric) continuous wavefunctions.

To arrive at the desired effective equation for the light field, we must consider that a cavity photon is converted into an exciton and, thus, evolves much more slowly. This is very closely related to the slow-light effect known from EIT propagation experiments \cite{PhysRevLett.84.5094} and, mathematically, it is found by solving all the coupled equations to first order in the slowly varying cavity field.
Neglecting interactions in Eq. \ref{eq:adiabatic_elim} we first iteratively solve the coherence for a single potential surface 
\begin{align*}
 \tilde{X}_{k}^{(1), \text{ single potential}}(r) = \frac{2 \Omega_{k}}{\Omega_{k}^{2}+\Gamma \Gamma_{k}} g \mathcal{E}(\vec{r})
\end{align*}
and then iteratively solve in the non-diagonal terms as done above giving
\begin{align*}
 \tilde{X}^{(1)}_{k}(\vec{r}) &= X^{(1)}_{k}(\vec{r}) - \frac{4 \Gamma \Omega_{k}}{(\Omega_{k}^{2} + \Gamma \Gamma_{k})^{2}} \underbrace{\left[ 1-(\Omega_{k}^{2} + \Gamma \Gamma_{k}) \sum_{k^{\prime} \neq k} 
 \frac{\Omega_{k^{\prime}}^{2}}{(\Omega_{k^{\prime}}^{2} + \Gamma \Gamma_{k^{\prime}})^{2}}\right]}_{\equiv F^{3}_{k}} g \partial_{t} \mathcal{E}(\vec{r}).
\end{align*}
The remaining terms do not contribute to any linear terms in $\partial_{t}\mathcal{E}$.
As done above, we substitute into Eq. \ref{eq:adiabatic_elim_basic} and find as the only modification
\begin{align}
 \tilde{X}_{g}^{(1)} = X_{g}^{(1)} - 4i \sum_{k} \frac{\Omega_{k}^{2}}{(\Omega_{k}^{2} + \Gamma \Gamma_{k})^{2}} \cdot F_{k}^{3} \cdot g \partial_{t} \mathcal{E}.
\end{align}
We define the (slow-light) factor $\nu = 1 + 4g^{2} \sum_{k} \frac{\Omega_{k}^{2}}{(\Omega_{k}^{2} + \Gamma \Gamma_{k})^{2}} \cdot F_{k}^{3}$, such that the in-cavity dynamics is given by
\begin{equation} \label{eq:cavity_equation}
\begin{aligned}
  \nu &\partial_{t} \mathcal{E}(\vec{r}) = i\frac{\hbar}{2 m_{ph}} \nabla^{2} \mathcal{E}(\vec{r}) - \frac{\Gamma_{\text{cav}}}{2} \mathcal{E}(\vec{r}) + \eta E^{\text{inc}}_{0}
  - ig \left( \chi^{(1)}
  + \int d^{2} r^{\prime} \chi^{(3)}(|\vec{r}-\vec{r}^{\prime}|) |\mathcal{E}(\vec{r}^{\prime})|^{2} \right) \mathcal{E}(\vec{r}).
\end{aligned}
\end{equation}
We define the nonlinear interaction $W(r) = \Re \left( \frac{g}{\nu} \chi^{(3)}(r) \right)$, which is shown in Fig. 1d (main text) and is the basis for Fig. 2 (main text). The corresponding imaginary component constitutes a nonlinear absorption term $\Gamma_{\text{nl}}(r) = \Im \left( \frac{g}{\nu} \chi^{(3)}(r) \right)$. The inset to Fig. 2a (main text) is calculated by solving Eq. (\ref{eq:cavity_equation}) for the (flat) cavity steady state $\partial_{t} \mathcal{E}(r) = 0$. 
The transmission is proportional to the ratio of the cavity steady state density and the input driving strength
\begin{equation}
\begin{aligned}
 T^{-1} &\propto \frac{(\eta E_0^{\text{inc}})}{|\mathcal{E}_0|^2} = 
 g^2 \left| \int d\vec{r} \chi^{(3)} (|\vec{r}|) \right|^2 |\mathcal{E}_0|^4 + \left( -\frac{\kappa}{2} + g\chi^{(1)}_I \right)^2 + \left( g\chi^{(1)}_R - \Delta_{\text{cav}} \right)^2 \\
 &+ 2g \left[ \int d\vec{r} \chi^{(3)}_R (|\vec{r}|) \cdot \left(g\chi^{(1)}_R - \Delta_{\text{cav}} \right) + 
 \int d \vec{r} \chi^{(3)}_I (|\vec{r}|) \cdot \left( -\frac{\kappa}{2} + g\chi^{(1)}_I \right) \right] |\mathcal{E}_0|^2
\end{aligned}
\end{equation}
and is evaluated as a function of the cavity frequency (via $\Delta_{\text{cav}}$) in the inset to Fig. 2a of the main text.
The point of maxial transmission is therefore given by
\begin{align}
 \Delta_{\text{cav}} = g\chi^{(1)}_R + g \int d\vec{r} \chi^{(3)}_R (|\vec{r}|) |\mathcal{E}_0|^2	
\end{align}
For typical parameters the slow light factor can be approximated by the real number $\nu \approx 4\frac{g^{2}}{\Omega^{2}}$, such that the nonlinear cavity shift can be expressed in terms of $\alpha = \int d^2 r W(r)$ as 
\begin{align}
 \delta_{\text{nl}} \approx \frac{4 \alpha g^2}{\Omega^2} |\mathcal{E}_0|^2
\end{align}
as given in the main text.

\subsection{Some Limiting Cases} \label{sec:limiting_cases}
To provide more insights of the complicated expressions of the nonlinear potential in Eq. (\ref{eq:chi3}), we analyze some of its features:
At large distances the photonic potential goes to zero because the dipole interaction vanishes. 
Mathematically, we can see this from Eqs. (\ref{eq:chi3}-\ref{eq:coupling_1}), where $c^{*}_{[kl], \mu} \rightarrow \delta_{[kl], \mu}$ and $U_{\mu}  \rightarrow -(\Delta_{k} + \Delta_{l}) $, leaving no total contribution.
Saturation as a single-particle effect cannot enter in the assumed limit of weak driving fields.
The other limit is the plateau at small distances, where the dipole interaction pushes the molecular states far away from their product values.
At even smaller distances, the states are mixed even more strongly, some re-enter the two-photon resonance.
First, we consider the case $V^{dd}>\bar{\gamma}$, i.e. the interaction shift is larger than the decay-broadened Rydberg line.
From the general equation we observe that for relatively small Rydberg decay and away from new resonances the terms $\Gamma_{k}\Gamma$ suppress contributions other than the resonant one. The nonlinear susceptibility thus reads
\begin{align} \label{eq:sigma_3_simple}
 \chi^{(3)} (r) &= \frac{8g^3\Omega^{4}}{(\Omega^{2} + \Gamma\bar{\gamma} ) \cdot |\Omega^{2} + \Gamma\bar{\gamma}|^{2}} \cdot \frac{2 U(r)}{\Omega^{2} +\bar{\gamma}\Gamma + i \Gamma U(r)},
\end{align}
where we named the only active Rabi coupling to the Rydberg potential $\Omega$. The nonlinear optical response, characterized by $W(r)$ and $\Gamma_{\text{nl}}(r)$, is
\begin{equation}
 \begin{aligned}
  W(r) = \frac{4g^2\Omega^2U(r) \cdot \left[ \bar{\gamma}^2 (\gamma^2 + 4 \Delta^2) + 2 (\gamma \bar{\gamma} + \Delta U(r)) \Omega^2 + \Omega^4 \right]}
  {|\Omega^{2} + \Gamma \bar{\gamma}|^{2} \cdot |\Omega^{2} + \bar{\gamma}\Gamma + i \Gamma U(r)|^{2}}
 \end{aligned}
\end{equation}
\begin{equation}
 \begin{aligned}
  \Gamma_{\text{nl}}(r) = -\frac{4g^2\Omega^2U^2(r) \cdot \left[ \gamma^2 \bar{\gamma} + 4 \bar{\gamma} \Delta^2 + \gamma \Omega^2 \right]}
  {|\Omega^{2} + \Gamma \bar{\gamma}|^{2} \cdot |\Omega^{2} + \bar{\gamma}\Gamma + i \Gamma U(r)|^{2}},
 \end{aligned}
\end{equation}
where we used the very accurate approximation $\nu = \frac{4g^{2}\Omega^2}{(\Omega^2 + \Gamma \bar{\gamma})^2}$ for simplicity.
First consider the plateau height in the case of small separations, i.e. $U \rightarrow \infty$
\begin{align} 
 W(0) &= \frac{8g^2\Delta\Omega^4} {|\Omega^{2} + \Gamma \bar{\gamma}|^{2} \cdot |\Gamma|^{2}} \label{eq:plateau_full1} \\
 \Gamma_{\text{nl}}(0) &= -\frac{4g^2\Omega^2 \left[ \gamma^2 \bar{\gamma} + 4 \bar{\gamma} \Delta^2 + \gamma \Omega^2 \right]}{|\Omega^{2} + \Gamma \bar{\gamma}|^{2} \cdot |\Gamma|^{2}} \label{eq:plateau_full2} 
\end{align}
\begin{figure}
\begin{center}
  \includegraphics[height=.6\textwidth]{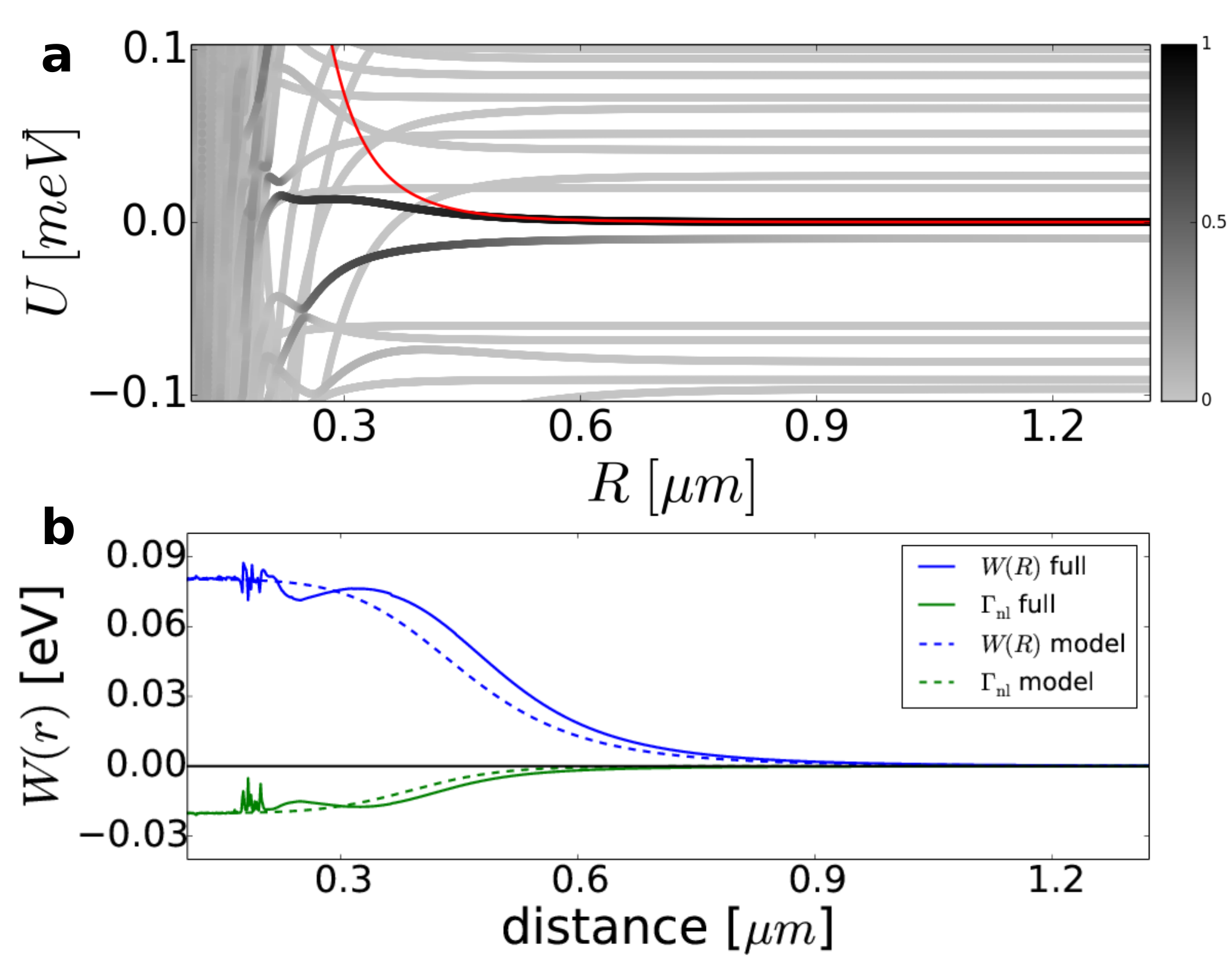}
\end{center}
\caption{a) Excitonic potential at $K^{+}, m=1$ and $n=10$, $\gamma = 300Ghz$, $\bar{\gamma} = 0.3Ghz$, $\Delta = 10\gamma$, $\Omega = 10 \delta_{\text{EIT}}$. Red line delineates $C_{6}/R^{6}$ fit at large distances. b) Optical potential calculated from full potential and single surface model. The plateau values agree. \label{fig:comp_single_multi_line}}
\end{figure}
Next, we compare our multilevel theory with the simplified version of a single van der Waals potential $U = C_{6}/R^{6}$ (Fig. (S\ref{fig:comp_single_multi_line}) a) by fitting $C_{6}$ at great distances and continuing the potential into smaller separations. At large distances the optical potentials of both cases tends to zero, at small distances both enter a plateau region whose height is entirely determined by the laser parameters. Remarkably, the plateau height is precisely the same
for the multilevel and the non-mixing single potential calculation (Fig. (S\ref{fig:comp_single_multi_line}) b). This is because the physical mechanism at the heart of the plateau is the loss of a coupling pair state, an effect which can either be brought about by a shift (van der Waals fit) or the combined action of shifting and state mixing (full potential). 
Since the plateau height is identical we use the simpler single potential relations to analyze its scaling relations.

For the special case $\bar{\gamma} = 0$ we find the familiar scaling relations in the dispersive regime $\Delta \gg \gamma$
\begin{align}
 W(0) \approx \frac{2g^2}{\Delta} \qquad \Gamma_{\text{nl}}(0) \approx -\frac{g^2\gamma}{\Delta^2}.
\end{align}
In this limit, we can get a dominant real part by increasing the detuning.
If $\bar{\gamma} \neq 0$, however, the plateau height scaling takes a different form and we can derive from Eqs. (\ref{eq:plateau_full1}-\ref{eq:plateau_full2}) that neglecting the Rydberg decay is legitimate if $\bar{\gamma} < \Omega^2/|\Gamma| = \delta_{EIT}$ and $\Delta > \gamma$ (disperive regime).

Another characteristic figure is the potential height $U^c = W(R_c)$ needed to form the potential, which we define via $W(r_c) = \frac{1}{2} W(0)$. This shift determines the potential's characteristic length scale $R_c$.
Even for the single potential surface this is a rather lengthy expression, we examine here the special case $\bar{\gamma} = 0$ finding
\begin{align}
 U^{c} = \frac{\Omega^2}{2\Delta} \cdot \frac{\pm \sqrt{ 1 + \left( \frac{\gamma}{2\Delta} \right)^2 + \left( \frac{\gamma}{2\Delta} \right)^4} - \left( \frac{\gamma}{2\Delta} \right)^2}{1 + \left( \frac{\gamma}{2\Delta} \right)^2} \rightarrow \pm \frac{\Omega^2}{2\Delta}.
\end{align}
In the dispersive regime we thus recover the simple physics outlined in the main text: The interactions must exceed leave the EIT window $\delta_{EIT} = \Omega^2/\Delta$ for a strong optical response.
Although the full potential rises less steeply than the van der Waals fit we observe that its optical plateau typically rises at greater distances.
This is, again, due to the loss of coupling strength thanks to state mixing.
Some more detailed potential features, such as local minima, are also not captured by the single potential model.
We conclude that while only the full response accounting for state mixing and resonances can capture all features,
the above estimates based on the van der Waals model provide a reliable approximation for most important aspects.

\section{Photon Correlations} \label{sect:phot_correlations}
In this section, we evaluate the quantum photon statistics of the cavity system described by Eqs. (\ref{eq:light_field}-\ref{eq:adiabatic_elim}).
We exploit that there is a natural separation of times scales. For realistic cavities (bad cavity limit) the light field will reach a steady state fast,
usually on timescale given by $\kappa^{-1}$. This is followed by a relaxation of the excitons' internal dynamics on a much longer timescale 
and finally the onset of transverse density currents in the light intensity driven by the kinetic energy operator on a yet longer timescale \cite{charmichael1988}.
Here, we focus on times shorter than necessary for the formation of intensity modulation, such that we can neglect the photonic kinetic energy term.
The formal solution of Eq. (\ref{eq:light_field}) reads, for coherent driving
\begin{equation}
\begin{aligned}
 \hat{\mathcal{E}}(\vec{r},t) &= e^{-\frac{\Gamma_{\text{cav}}\Gamma + 4g^{2}}{2\Gamma}t}\hat{\mathcal{E}}(\vec{r},0) 
 + \int_{0}^{t}e^{-\frac{\Gamma_{\text{cav}}\Gamma + 4g^{2}}{2\Gamma}(t-t^{\prime})} \left( \frac{g}{\Gamma} \sum_{i} \Omega_{i} \gry{i}(\vec{r}, t^{\prime})  + \eta E^{\text{in}}(\vec{r}) \right) dt^{\prime}.
\end{aligned}
\end{equation}
Since the cavity is initially empty we drop the term proportional to $\hat{\mathcal{E}}(r,0)$.
Because of the timescale separation it now is possible to solve the integral in the Markov approximation. 
We first formulate the integral with a memory kernel (which does not change the limits of integration), $\tau = t-t^{\prime}$
\begin{align}
 \int_{0}^{t}e^{-\frac{\Gamma_{\text{cav}}\Gamma + 4g^{2}}{2\Gamma}\tau} \left( \frac{g}{\Gamma} \sum_{i} \Omega_{i} \gry{i}(\vec{r}, t-\tau)  + \eta E^{\text{in}}(\vec{r}) \right) d\tau
\end{align}
and replace $\gry{i}(\vec{r}, t-\tau) \rightarrow \gry{i}(\vec{r}, t)$ under the integral. Carrying out the integral, we find the solution after the very short initial time, i.e. for $t \gg \kappa^{-1}$
\begin{align}
 \hat{\mathcal{E}}(\vec{r},t) &= \frac{2g}{\Gamma_{c}\Gamma + 4g^{2}} \sum_{i} \Omega_{i} \hat{X}_{i}(\vec{r},t) + \frac{2\Gamma \eta}{\Gamma_{c}\Gamma + 4g^{2}} E^{\text{in}}(\vec{r}).
\end{align}
Plugging this adiabtic solution into Eq. (\ref{eq:adiabatic_elim}), we are left with an equation of motion only in the excitonic degrees of freedom
\begin{equation}\label{eq:quantum_all_states}
 \begin{aligned}
 \partial_{t} \hat{X}_{k}(\vec{r}) &= \frac{\Omega_{k}}{\Gamma} \left( - \frac{1}{2} + \frac{2g^{2}}{\Gamma_{c}\Gamma + 4g^2} \right) \sum_{i} \Omega_{i} \hat{X}_{i}(\vec{r})  - \frac{\Gamma_{k}}{2} \hat{X}_{k}(\vec{r}) + \frac{2 \eta g \Omega_{k}}{\Gamma_{c}\Gamma + 4g^{2}} E^{\text{in}}(\vec{r}) \\
  &- i\sum_{i \leq j, i^{\prime}} \int d^{2}r^{\prime} \hat{X}^{\dagger}_{i^{\prime}}(\vec{r}^{\prime}) \frac{V^{dd}_{i^{\prime}k, ij}(|\vec{r}^{\prime} - \vec{r}|)}{\hbar} \hat{X}_{i}(\vec{r}^{\prime}) \hat{X}_{j}(\vec{r}).
\end{aligned}
\end{equation}
We note that Eq. (\ref{eq:quantum_all_states}) is formally equivalent to Eq. (\ref{eq:adiabatic_elim}), although the former includes quantum correlations
between excitons and the cavity field.
Thanks to this formal identity we can employ the same solution strategy as above, replacing only certain variables.
As demonstrated in Sec. (\ref{sec:limiting_cases}), reduction to a system with one primary Rydberg level (created by $\hat{X}_{s}^{\dagger}$) interacting via van der Waals interactions $V(R) = \hbar \omega_{0} /(R/R_c)^{6}$ produces
a good approximation to the full solution. For simplicity of notation and computation we will consider this simplified model in the following
\begin{align}\label{eq:field_simplified}
  \eps &= \underbrace{\frac{2g\Omega}{\Gamma_{c}\Gamma + 4g^{2}}}_{\equiv \gamma} \hat{X}_{s}(\vec{r}) + \underbrace{\frac{2\Gamma \eta}{\Gamma_{c}\Gamma + 4g^{2}}}_{\equiv \delta} E^{\text{in}}(\vec{r})
\end{align}
\begin{equation}
 \begin{aligned}
  \partial_{t} \hat{X}_{s}(\vec{r}) &= \underbrace{\left( \frac{\Omega^{2}}{\Gamma} \left( - \frac{1}{2} + \frac{2g^{2}}{\Gamma_{c}\Gamma + 4g^2} \right) - \frac{\Gamma_{s}}{2} \right)}_{\equiv \alpha} \hat{X}_{s}(\vec{r}) + \underbrace{\frac{2 \eta g \Omega}{\Gamma_{c}\Gamma + 4g^{2}}}_{\equiv \beta} E^{\text{in}}(\vec{r})
  - i \int d\vec{r}^{\prime} \hat{X}^{\dagger}_{s}(\vec{r}^{\prime}) \frac{\omega_{0} R_c^{6}}{|\vec{r}-\vec{r}^{\prime}|^6} \hat{X}_{s}(\vec{r}^{\prime}) \hat{X}_{s}(\vec{r}).
 \end{aligned}
\end{equation}
The photon statistics $g^{(2)}(0)$ requires spatially resolved information on the photonic operators, in particular on the correlation function
\begin{align}
 \Braket{\hat{\mathcal{E}}^{\dagger}(\vec{r}_1)\hat{\mathcal{E}}^{\dagger}(\vec{r}_2)\hat{\mathcal{E}}(\vec{r}_1)\hat{\mathcal{E}}(\vec{r}_2)},
\end{align}
which can be expressed in terms of excitonic operators by virtue of Eq. (\ref{eq:field_simplified})
\begin{equation}
 \begin{aligned}
  |\gamma|^{4} \hat{X}_{s}^{\dagger}(\vec{r}_1)\hat{X}_{s}^{\dagger}(\vec{r}_2)\hat{X}_{s}(\vec{r}_1)\hat{X}_{s}(\vec{r}_2) + |\gamma|^{2}\gamma^{*}\delta E^{\text{in}}(\vec{r}_2) \hat{X}_{s}^{\dagger}(\vec{r}_1) \hat{X}_{s}^{\dagger}(\vec{r}_2) \hat{X}_{s}(\vec{r}_1) \\
  +|\gamma|^{2} \delta^{*}\gamma E^{\text{in}*}(\vec{r}_2) \hat{X}_{s}^{\dagger}(\vec{r}_1) \hat{X}_{s}(\vec{r}_1)\hat{X}_{s}(\vec{r}_2) + |\gamma|^{2} |\delta|^{2} E^{\text{in}}(\vec{r}_2)E^{\text{in}*}(\vec{r}_2) \hat{X}_{s}^{\dagger}(\vec{r}_1) \hat{X}_{s}(\vec{r}_1) + ...
 \end{aligned}
\end{equation}
Out of the 16 terms in this expansion all but the very first term have already been evaluated in Sec. (\ref{sec:nonlinear_optical_response}).
This first term describes the excitonic blockade or the expectation value of finding Rydberg-excited excitons in positions $\vec{r}_1$ and $\vec{r}_2$
and is, in fact, a linear combination of the other (known) two-particle terms.
For completeness, all necessary terms are given below
\begin{align}
 \Braket{\hat{X}_{s}^{\dagger}(\vec{r}_1)\hat{X}_{s}(\vec{r}_1)} &= \frac{-\beta E^{\text{in}}(\vec{r}_1)\Braket{\hat{X}^{\dagger}_{s}(\vec{r}_1)} - \beta^{*} E^{\text{in}*}(\vec{r}_1) \Braket{\hat{X}_{s}(\vec{r}_1)}}{\alpha + \alpha^{*}}
\end{align}
\begin{align}
 \Braket{\hat{X}_{s}(\vec{r}_1)\hat{X}_{s}(\vec{r}_2)} &= \frac{\beta \left( E^{\text{in}}(\vec{r}_1) \Braket{\hat{X}_{s}^{\dagger}(\vec{r}_2)} + E^{\text{in}}(\vec{r}_2) \Braket{\hat{X}_{s}(\vec{r}_1)} \right)}{-2\alpha + i V(|\vec{r}_1 - \vec{r}_2|)} \\
 \Braket{\hat{X}_{s}^{\dagger}(\vec{r}_2)\hat{X}_{s}(\vec{r}_1)} &= \frac{-\beta E^{\text{in}}(\vec{r}_1)\Braket{\hat{X}^{\dagger}_{s}(\vec{r}_2)} - \beta^{*} E^{\text{in}*}(\vec{r}_2) \Braket{\hat{X}_{s}(\vec{r}_1)}}{\alpha + \alpha^{*}}
\end{align}
\begin{equation}
\begin{aligned}
 \Braket{\hat{X}_{s}^{\dagger}(\vec{r}_2)\hat{X}_{s}(\vec{r}_1)\hat{X}_{s}(\vec{r}_2)} = \frac{1}{-2\alpha - \alpha^{*} + i V(|\vec{r}_1 - \vec{r}_{2}|)} 
 \left[ \beta E^{\text{in}}(\vec{r}_1) \Braket{\hat{X}_{s}^{\dagger}(\vec{r}_2) \hat{X}_{s}(\vec{r}_2)} \right. \\
 \left. +\beta^{*} E^{\text{in}*}(\vec{r}_2) \Braket{\hat{X}_{s}(\vec{r}_1)\hat{X}_{s}(\vec{r}_2)} \right. \\
 \left. + \beta E^{\text{in}}(\vec{r}_2) \Braket{\hat{X}_{s}^{\dagger}(\vec{r}_2) \hat{X}_{s}(\vec{r}_1)} \right]
\end{aligned}
\end{equation}
\begin{equation}
\begin{aligned}
 \Braket{\hat{X}_{s}^{\dagger}(\vec{r}_1)\hat{X}_{s}^{\dagger}(\vec{r}_2)\hat{X}_{s}(\vec{r}_1)\hat{X}_{s}(\vec{r}_2)} = \frac{1}{2(\alpha + \alpha^{*})} \left[ -\beta^{*}E^{\text{in}*}(\vec{r}_2)\Braket{\hat{X}_{s}^{\dagger}(\vec{r}_2)\hat{X}_{s}(\vec{r}_1)\hat{X}_{s}(\vec{r}_2)} \right. \\
 \left. -\beta E^{\text{in}}(\vec{r}_1) \Braket{\hat{X}_{s}^{\dagger}(\vec{r}_1)\hat{X}_{s}^{\dagger}(\vec{r}_2) \hat{X}_{s}(\vec{r}_2)}  \right. \\
 \left. -\beta^{*} E^{\text{in}*}(\vec{r}_1) \Braket{\hat{X}_{s}^{\dagger}(\vec{r}_1)\hat{X}_{s}(\vec{r}_1) \hat{X}_{s}(\vec{r}_2)} \right. \\
 \left. -\beta E^{\text{in}}(r_2) \Braket{\hat{X}_{s}^{\dagger}(\vec{r}_1)\hat{X}_{s}^{\dagger}(\vec{r}_2) \hat{X}_{s}(\vec{r}_1)} \right].
\end{aligned}
\end{equation}
This system of equations leads to a nonlinear equation in the single-particle correlator.
We solve this equation for weak driving fields by plugging in the (approximate) noninteracting solution.
It turns out that the external driving field can be factored out of all terms and cancels with the denominator
\begin{equation}
\begin{aligned}
 &h^{(2)}(\vec{r}_1,\vec{r}_2)\equiv \frac{\Braket{\hat{\mathcal{E}}^{\dagger}(\vec{r}_1)\hat{\mathcal{E}}^{\dagger}(\vec{r}_2)\hat{\mathcal{E}}(\vec{r}_1)\hat{\mathcal{E}}(\vec{r}_2)}}{\Braket{\hat{\mathcal{E}}^{\dagger}(\vec{r}_1)\hat{\mathcal{E}}(\vec{r}_1)}\Braket{\hat{\mathcal{E}}^{\dagger}(\vec{r}_2)\hat{\mathcal{E}}(\vec{r}_2)}} \\
 &= \frac{1}{|-\gamma \frac{\beta}{\alpha} + \delta|^{4}} \left( 4 \left| \frac{\beta \gamma \delta}{\alpha} \right|^{2} + |\delta|^{4} \right. \\
 &\left. + 4\Re \left[ \left( \frac{\beta^{*}|\beta/\alpha|^{2}}{-2\alpha^{*}-\alpha-iV(|\vec{r}_1-\vec{r}_2|)} + \frac{\beta^{*2}/\alpha^{*}}{2\alpha^{*}+iV(|\vec{r}_1-\vec{r}_2|)} \right) \cdot \left( -\frac{\beta |\gamma|^{4}}{\alpha+\alpha^{*}} + 2|\gamma|^{2}\gamma^{*}\delta \right) \right. \right. \\
 &\left. \left. (\gamma^{*}\delta)^{2} \frac{\beta^{*2}/\alpha^{*}}{2\alpha^{*}+iV(r_1-r_2)} - \frac{\gamma^{*}\delta|\delta|^{2} \beta^{*}}{\alpha^{*}}\right] \right).
\end{aligned}
\end{equation}
The limit of strong interactions reached at short distances reads
\begin{align}
 h^{(2)}(0) = \frac{4 \left| \frac{\beta \gamma \delta}{\alpha} \right|^{2} + |\delta|^{4} - 4 \left| \frac{\delta}{\alpha} \right|^{2} \Re \left[ \alpha \beta^{*} \gamma^{*} \delta \right]}{\left|-\gamma \frac{\beta}{\alpha} + \delta \right|^{4}}.
\end{align}
For illustration, we consider a setup in which the incoming beam is flat on a disk of diameter $d$ and zero elsewhere. This could be realized by placing a mask
on top of the TMDC layer, where we restrict our attention to a single hole behind which a detector counts all exiting photons without spatial resolution.
Any matrix element is symmetric w.r.t. the interchange $\vec{r}_1 \leftrightarrow \vec{r}_2$, which faciliates the numerics and leads to $h^{(2)}(\vec{r}_1,\vec{r}_2) = h^{(2)}(|\vec{r}_1 - \vec{r}_2|)$.
The measured quantity would be
\begin{equation}
\begin{aligned}
 \bar{g}^{(2)}(\tau = 0) &= \int d\vec{r}_1 \int d\vec{r}_2 h^{(2)}(|\vec{r}_1 - \vec{r}_2|) = \int_{\vec{r}_1, \vec{r}_2 \in \text{disk}} d\vec{r} d\vec{R} h^{(2)}(|\vec{r}|) \\
   &= 2 \pi \int_{0}^{d} dr \ r \cdot 2 \left[ \left( \frac{d}{2} \right)^{2}\arccos \left( \frac{r}{d} \right) - \frac{r}{2} \sqrt{\left( \frac{d}{2} \right)^2 - \left( \frac{r}{2} \right)^{2}} \right] \cdot h^{(2)}(r)
\end{aligned}
\end{equation}
where we moved to center of mass coordinates $\vec{r} = \vec{r}_2 - \vec{r}_1$, $\vec{R} = (\vec{r}_1 + \vec{r}_2)/2$ (Jacobi-determinant is 1). In the second line the inner integral depends on the outer and
maps out two circle segments (therefore a factor of 2) of height $\left( \frac{d}{2} \right) - r/2$ each, the outer angular integral can then be performed explicitly.
The limits of integration are $[0;d]$ because the relative coordinate can be as long as the detector's diameter.
It makes sense to normalize to the detector area and define
\begin{equation}
\begin{aligned}
 g^{(2)}(\tau = 0) &= \frac{\bar{g}^{(2)}(\tau = 0)}{\pi^{2}\left(\frac{d}{2} \right)^4}
 = \frac{64}{\pi d} \int_{0}^{d} dr \ r \cdot \left[ \left( \frac{d}{2} \right)^{2}\arccos \left( \frac{r}{d} \right) - \frac{r}{2} \sqrt{\left(\frac{d}{2} \right)^2 - \left( \frac{r}{2} \right)^{2}} \right] \cdot g^{(2)}(r)
\end{aligned}
\end{equation}
with $g^{(2)}(\tau = 0) = 1$ for the noninteracting case. We then define define the distance $R_{\mathrm{b}}^{(\mathrm{ph})}$ as the spot diameter at which the function $g^{(0)}(0)$ has dropped to $\frac{1}{2}$. The analogous definition applies to $R_{\mathrm{b}}^{(\mathrm{X})}$, which measures the excitonic rather than the photonic correlation on the basis of $\Braket{\hat{X}_{s}^{\dagger}(\vec{r}_1)\hat{X}_{s}^{\dagger}(\vec{r}_2)\hat{X}_{s}(\vec{r}_1)\hat{X}_{s}(\vec{r}_2)}$. The results of this calculation are shown in Fig. 3 of the main text.

\end{document}